\documentclass[prl,twocolumn]{revtex4}

\usepackage{graphicx}

\begin{document}

\title{Ground State Properties of Cold Bosonic Atoms At Large Scattering Lengths}
\author{Jun Liang Song and  Fei Zhou}
\affiliation{Department of Physics and Astronomy, The University of British Columbia, Vancouver, B. C., Canada V6T1Z1}
\date{{\small \today}}

\begin{abstract}
In this Letter, we study bosonic atoms at large scattering lengths
using a variational method where the condensate amplitude is a variational
parameter. We further examine momentum distribution functions,
chemical potentials, the speed of sound,
and spatial density profiles of cold bosonic atoms in a trap in this
limit. The later two
properties turn out to bear similarities of those of Fermi gases.
The estimates obtained here are applicable near Feshbach resonances, particularly when
the fraction of atoms forming
three-body structures is small and can be tested in future cold atom
experiments.
\end{abstract}
\maketitle

Bose-Einstein condensates(BECs) near Feshbach resonances have been one of
the most exciting ultracold systems
studied so far in
experiments\cite{Sackett99,Donley01,Roberts01,Kagan97,Inouye98,Cornish00,Claussen02,Papp08, Pollack09}.
On one side of the resonances where the scattering lengths are negative,
fascinating collapse-growth cycles due to thermal clouds and spectacular controlled collapsing-exploding
dynamics have been observed\cite{Sackett99,Donley01,Roberts01} and studied
theoretically\cite{Kagan97}.
On the other side, towards the resonances where scattering lengths are positive,
strongly repulsive ultracold bosonic atoms and their intriguing properties
have been explored\cite{Inouye98,Cornish00,Claussen02,Papp08}.
Despite the reduced lifetime of the cold gases in this limit due to enhanced three-body recombination,
quite remarkable progress has been made to probe interactions between atoms.
Recently, pursuit in this direction has been
revived, and more vigorous efforts have been made\cite{Papp08, Pollack09}.
Our theoretical studies in this Letter are mainly motivated by these
experiments.
Bose gases at large positive scattering lengths have
been a horrendously challenging topic in theoretical physics for
more than half a
century\cite{Bogoliubov47,Lee57a,Lee57b,Beliaev58,Nozieres90}.
The standard low density expansion that works quite well for dilute gases
is not applicable when the scattering length $a$ is comparable to or even
larger than the mean interparticle distance $d$.
Here we suggest a variational approach which takes into account
{\em two-body} correlations
and can be extended to the limit of a large
positive scattering length.
We further apply this approach
to estimate various fundamental properties of cold bosonic atoms near
Feshbach resonances,
particularly when the fraction of atoms forming three-body structures is small.
Unique features in the
momentum distribution function, chemical potential, speed of sound and
the cold atom density profile in a trap can be potentially probed
in experiments.

Cold bosonic atoms at large scattering lengths were
also previously addressed in a few inspiring theoretical
papers\cite{Pethick02,Braaten02}.
Cowell {\em et al.} estimated chemical potentials and condensate
fractions by
employing distinctly different Jastrow
wavefunctions\cite{Pethick02}.
There are a few interconnected differences between their results and ours.
First, while the physics at distances much shorter than the mean
interparticle distance $d$ is described quite accurately by the
Jastrow wave functions, basic aspects of the long wavelength physics are not expected to be well captured.
On the other hand, our wave function is constructed under a
constraint in Eq.(\ref{vac})
and captures essential features of low energy
collective properties of BECs.
For instance, the momentum distribution function $n_{\bf k}$
has a $\frac{1}{k}$ divergence near $k=0$ for all scattering
lengths, and at short distances our
wave function is almost identical to the
solution to the Schr\"odinger equation for two interacting atoms.
Second, since the contribution to the depletion fraction, or the
fraction of atoms occupying nonzero momentum states,
is mainly from low energy states,
we expect that our results are more reliable.
In fact, we find that the depletion fraction reaches a constant value
of about $0.5$ near resonances.
On the contrary, the condensate fraction estimated in
Ref.\cite{Pethick02}
quickly
reaches zero when the scattering length $a$ becomes comparable to $d$,
suggesting that atoms could be completely depleted from the zero momentum
state and there should be an unexpected quantum phase transition at a
finite
scattering length.
Third, chemical potentials near resonances estimated there
appear to be bigger than the values obtained in our calculations.
This seems to imply that the trial wave functions adopted here
should be an energetically better candidate for ground states.

The trial wave function in Eq.(\ref{wf}) effectively encodes
two-body correlations. To include high-order correlations such as three-body
effects, a much more sophistic ansatz is needed.
A nontrivial role of three-body
interactions was previously appreciated by
Braaten {\em et al.} in Ref.\cite{Braaten02}, where
the effects on BECs were estimated in the limit of a small scattering length.
Although there was no definite evidence of Efimov trimers in
BECs of sodium or rubidium atoms studied in
Ref.\cite{Inouye98,Cornish00,Claussen02,Papp08},
an earlier experiment on relaxation rates of cesium atoms did show, as a
precursor of two-body resonances,
additional structures which had been attributed to
Efimov states\cite{Efimov70,Kraemer06,Esry99,Braaten03}.
More efforts are to be made to
understand the nature of BECs in
this limit
and the approach proposed below is a baby step towards this direction.
Our results are valid when the three-body correlations
induced by Efimov trimers are not dominating.
The question of whether the emergence of Efimov trimers
introduces distinct modulations to the
scaling functions discussed below, or mainly sets the
lifetime of BECs, represents an exciting new direction that is worth pursuing.

Moreover, our scaling hypothesis works best when
the typical range of interactions $r_0$ is much
less than the inter-particle distance $d$. When the density increases,
deviations from the scaling behaviors
become substantial,
and the scaling functions
proposed below are no longer sufficient for
characterizing BECs.
Eventually, a quantum gas
might undergo a transition
to a dense liquid phase when $r_0$ becomes comparable to $d$.
For cold atoms, this fortunately only occurs at a density which
is not experimentally accessible because of severe trap losses.

We consider bosonic atoms that interact with a short range potential of range $r_0$ and scatter at
two-body scattering lengths $a(>0)$. For BECs with a number density $\rho_0$,
assuming two-body effects are dominating, we can generally express the momentum distribution function $n_{\bf k}$ and
the chemical potential $\mu$ in terms of dimensionless functions $f$ and $h$, i.e., $n_{\bf k}=f(kd,\frac{a}{d},
\frac{r_0}{d})$, $\mu=\epsilon_F
h(\frac{a}{d},\frac{r_0}{d})$; and $d=(\frac{3}{4 \pi \rho_0})^{1/3}$,
$\epsilon_F=\frac{(6\pi^2\rho_0)^{2/3}}{2m}$.
For short range interactions, $r_0$ is much smaller than
the mean inter-particle distance $d$
so that we approximate $\frac{r_0}{d}$ to be zero but $a$
can vary over a range from much smaller than $d$ to much bigger than $d$.
Function $f$ and $h$ thus depend only on two {\em dimensionless} variables
$x=kd$ and $y=\frac{a}{d}$
and are reduced to two scaling functions
$f(x,y)$ and $h(y)$, respectively.
The functional form of $f(x,y)$ and $h(y)$ proposed in this way does not depend on details of interaction
potentials or number densities or scattering lengths and is universal; $f$ and $h$ characterize basic properties of
BECs.
Note that, as illustrated below, the Fermi energy $\epsilon_F$
that is normally defined for a Fermi gas with the same number density
$\rho_0$
turns out to be the only relevant energy scale for BECs near
resonances.

When $a$ is much smaller than $d$,
these functions can be obtained by using the standard
mean field\cite{Beliaev58,Lee57a,Lee57b,Nozieres90}.
Indeed, in the
dilute limit when $y$ is much less than unity one can verify
that
\begin{eqnarray}
&& f(x,y)=\frac{1}{2}
\left(\frac{x^2+ 6 y}{\sqrt{x^2(x^2+ 12 y)}}-1 \right),
\nonumber \\
&& h(y)=(\frac{32}{3\pi^2})^{1/3} y, ~ ~ g(y)=\frac{4}{\sqrt{3} \pi}y^{3/2},
\label{distribution}
\end{eqnarray}
where we also introduce $g(y)$ for the depletion fraction.
$f(x,y)$ is divergent as $\sqrt{y}/x$
when $x$ or momentum $k$ approaches zero; this behavior is an indication of gapless soundlike
collective excitations in BECs. Furthermore,
that $f(x,y)$ decays as $y^2/x^4$
in the large-$x$ or large-$k$ limit
reflects the free particle nature of high energy excitations. For cold atoms at large scattering lengths, $y$
is substantial and the form of
$f$ and $h$ functions remains to be understood.
In the following we are going to investigate these scaling
functions in the limit when $a$(or $y$)
becomes comparable to or bigger than $d$(or $1$).

To quantitatively study $f$ and $h$ functions in the limit of a large scattering length,
we adopt a variational approach to BECs.
In this method, $c_0$, the condensate amplitude
and $g_{\bf k}, {\bf k}\neq 0$, the pairing amplitude that
is related to the occupation number of atoms in a state of momentum ${\bf
k}$, are variational parameters.
We then minimize the energy with respect to
$g_{\bf k}$ and $c_0$ but
with the total number of atoms $N_T$ fixed.

To introduce trial wave functions which are viable in both small and large scattering-length
limits, we {\em require} that at any given scattering length the
ground state should be a vacuum
of Bogoliubov quasi-particles and is
annihilated by a set of quasiparticle operators
\begin{eqnarray}
\left(\frac{1}{\sqrt{1-|g_{\bf k}|^2}}\hat{a}_{\bf k} - \frac{g_{\bf k}}{\sqrt{1-|g_{\bf k}|^2}} \hat{a}^{\dagger}_{\bf -k} \right)
|\textrm{g.s.}\rangle = 0.
\label{vac}
\end{eqnarray}
Here $\hat{a}_{\bf k}$ ($\hat{a}^\dagger_{\bf k}$) is
an annihilation (creation) operator for an atom with momentum
${\bf k}$. Detailed
structures of the quasi-particle operators
are specified by real variables $g_{\bf k}$ and will be determined
variationally below.
The ansatz that satisfies Eq.(\ref{vac}) can be written as
\begin{eqnarray} \label{wf}
|\textrm{g.s.}\rangle =
{\cal A}^{-1/2} \exp{\left( c_0 \hat{a}^{\dagger}_0\right)}
\prod_{\bf k\cdot \hat{z} >0}
\exp\left(g_{\bf k} \hat{a}^\dagger_{\bf k} \hat{a}^\dagger_{\bf -k}\right)|0\rangle.
\end{eqnarray}
Here ${\cal A}$ is the normalization factor.
Again $c_0$ is the condensation amplitude and $g_{\bf k}$ is the pairing amplitude with $|g_{\bf k}|<1$;
for ground states, we further assume $g_{\bf -k} = g_{\bf k}$.
This trial wave function encodes two-body correlations but not three-body ones.
Similar wave functions have been used to study
pair condensates of attractive bosons\cite{Nozieres82}.
$n_{\bf k}$, the occupation number of atoms with momentum ${\bf k}$, is a simple function of $g_{\bf k}$
\begin{eqnarray}
n_{\bf k}=\langle \hat{a}^\dagger_{\bf k} \hat{a}^{}_{\bf k} \rangle
&=& \frac{|g_{\bf k}|^2}{1-|g_{\bf k}|^2}.
\label{momfunct}
\end{eqnarray}

The Hamiltonian of cold bosons is
\begin{eqnarray}
H = \sum_{\bf k} \epsilon_{\bf k} \hat{a}^\dagger_{\bf k}\hat{a}^{}_{\bf k} +
\frac{1}{2} \sum_{\bf k_1, k_2, q}
\hat{a}^\dagger_{{\bf k_1}+{\bf q}}\hat{a}^\dagger_{{\bf k_2}-{\bf q}} U({\bf q})
\hat{a}_{\bf k_1}\hat{a}_{\bf k_2},
\end{eqnarray}
$U({\bf q}) =\frac{1}{\Omega} \int d^3r U({\bf r}) \exp(i {\bf q}\cdot {\bf r})$
is a two-body interaction potential,
and $\Omega$ is the volume of the system.
The total energy $E_T$ of the trial state $|g.s.>$ is evaluated to be
\begin{eqnarray} \label{eqn:energy}
E_T &=& \sum_{\bf k} \epsilon_{\bf k} \frac{|g_{\bf k}|^2}{1-|g_{\bf k}|^2}
+\frac{U(0)}{2} |c_0|^4 \nonumber \\
&+&  \sum_{\bf k,q\neq0}  \frac{U({\bf k-q}) + U(0)}{2} \frac{|g_{\bf k}|^2}{1-|g_{\bf k}|^2}  \frac{|g_{\bf q}|^2}{1-|g_{\bf q}|^2}\nonumber \\
&+&  \sum_{\bf k,q\neq0} \frac{U({\bf k-q})}{2} \frac{g^*_{\bf q}}{1-|g_{\bf q}|^2} \frac{g_{\bf k}}{1-|g_{\bf k}|^2} \nonumber \\
&+&  \sum_{\bf q\neq0}\frac{U({\bf q}) + U(0)}{2} \frac{2|g_{\bf q}|^2}{1-|g_{\bf q}|^2} |c_0|^2  \nonumber \\
&+& \sum_{\bf q\neq0} \frac{U({\bf q})}{2} \frac{c_0^2 g^*_{\bf q} + c_0^{*2} g_{\bf q}}{1-|g_{\bf q}|^2}.
\end{eqnarray}
To facilitate discussions on large scattering lengths,
we assume that the interaction potential is a square well
one, $U(r)=-U$ when $r<r_0$ but otherwise is zero.
The corresponding $s$-wave scattering length is
$a = r_0-\tan\left(\sqrt{mU}r_0\right)/\sqrt{mU}$.
We choose the depth of the potential $U$ to be $ \pi/2 < \sqrt{mU}r_0 < \pi$
so that $r_0 < a < \infty$.

\begin{figure}
\includegraphics[width=\columnwidth]{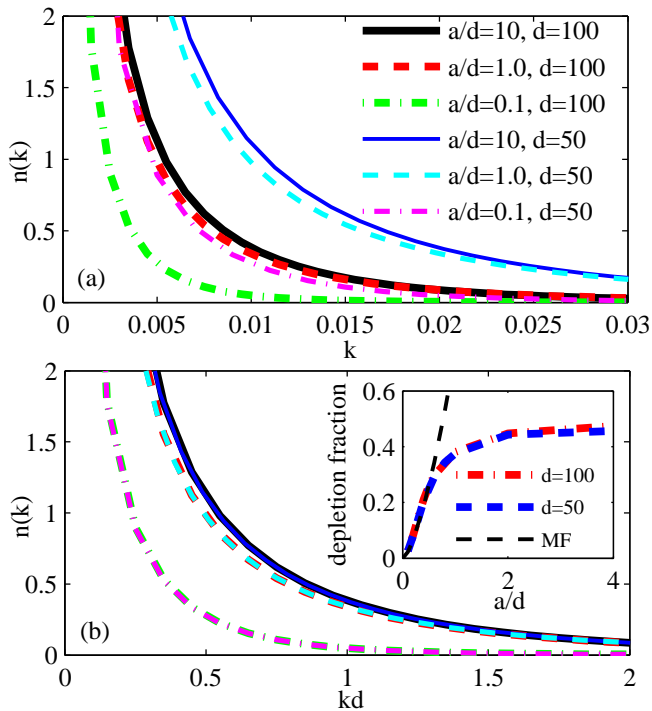}
\caption{(Color online) The momentum distribution function $n(k)$ for
different interparticle distance $d$ and
scattering length $a$ [in units of $r_0$,
the range of interaction and in a, $k$ is in units of $\hbar/r_0$ ].
In all cases, $n(k)$ has a desired $1/k$ divergence when $k$ approaches
zero.
In b), $n(k)$ functions for different $d$ but with the same value of $y=\frac{a}{d}$
are further shown to collapse to a {\em single} scaling function when plotted against $x=kd$.
The resultant three curves are for $f(x,y)$ with $y=0.1, 1$ and $10$ (from bottom to top).
Depletion fraction $g(y)$
is plotted in the inset; the mean field(MF) $g(y)$
in Eq.(\ref{distribution}) is also shown as a reference.}
\label{fig1}
\end{figure}

To obtain ground states,
we minimize the total energy in Eq.(\ref{eqn:energy}) with respect to
parameters $g_{\bf k}$ and
$c_0$, subject to a constraint that the total number $N_T$ is fixed,
\begin{eqnarray}
N_T =|c_0|^2+\sum_{{\bf k} \neq 0 }\frac{|g_{\bf k}|^2}{1-|g_{\bf k}|^2}.
\end{eqnarray}
When the potential is weakly repulsive, we verify that the minimization
does lead to the standard results for weakly interacting BECs, i.e.
Eq.(\ref{distribution}). For attractive potentials introduced above,
the minimization is carried out numerically.
When scattering lengths are positive,
one of the energy minima turns out to be
a collection of molecules as expected from a two-body consideration; in these molecular states, the condensate
amplitude is found to be zero and $|g_{\bf k}|$ is less than unity for
all ${\bf k}$. To understand BECs of scattering atoms
in open or nonmolecular channels that are most relevant to experiments
on cold atoms, we project away the molecular states and minimize the energy in the subspace of
scattering channels. This is achieved by imposing a projection constraint on $g_{\bf k}$,
$\sum_{\bf k} g^{mol}_{\bf k} g^*_{\bf k}=0$,
where $g_{\bf k}^{mol}$ are the calculated values of $g_{\bf k}$ for
molecular states.
This vanishing inner product between molecular states and
states of atoms
effectively projects out a desired subspace of open channels.

\begin{figure}
\includegraphics[width=\columnwidth]{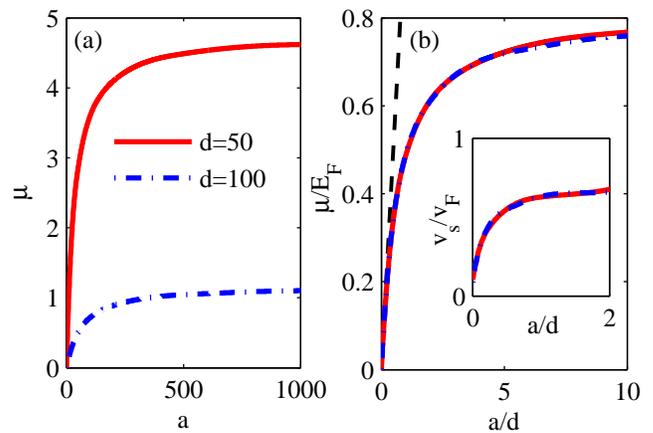}
\caption{(Color online) Chemical potentials $\mu$ and sound velocities
$v_s$ as a function of scattering length.
a) is for $\mu$ (in units of $1/2m d^2$, $d=50$ versus $a$
($d$ and $a$ are in units of $r_0$);
in b) we further plot $\mu$ in units of the Fermi energy
$\epsilon_F$
as a function of $y=a/d$ and
illustrate two plots in a) collapse into a single scaling curve.
The resultant plot defines the scaling function
$h(y)(=\mu/\epsilon_F)$ for an arbitrary $y$.
The dashed line is for $h(y)$ in Eq.(\ref{distribution}).
Shown in the inset is $v_s$ (in units of the Fermi velocity $v_F$ ) versus $a/d$.}
\label{fig2}
\end{figure}

Below we present results for BECs with
different densities and scattering lengths.
The minimization algorithm does converge leading to a ground state in the
subspace when we set $g_{\bf k} + 1$ to be proportional to
$k$ in the close vicinity of $k=0$\cite{simulations}.
We further find that $g_{\bf k}$ decays as $\frac{1}{k^2}$
in the large-$k$ limit for all scattering lengths. Following the relation
between $g_{\bf k}$ and $n_{\bf k}$ in Eq.(\ref{momfunct}), one then obtains the asymptotics of $n_{\bf k}$ in
both large-$k$ and small-$k$ limits. The characteristics in these two limits are robust
and, when the scattering length $a$ is tuned,
remain the same as those in Eq.(\ref{distribution}).
However, the crossover energy between these two limits, which is approximately
the chemical potential, strongly
depends on the scattering lengths or densities (see Fig.\ref{fig1}).
When plotted against $x=kd$, data for $n_{\bf k}$ or $n(k)$
calculated for different densities and
scattering lengths all collapse to a single set of curves
which correspond to $f(x,y)$ for different
$y=\frac{a}{d}$.
Furthermore, we observe that the function $n(k)=f(x,y)$ quickly
approaches $f_\infty (x)$ when
$y$ exceeds unity.
Using the momentum distribution function, we also estimate $g(y)$, the fraction of atoms
that are depleted from the zero momentum state;
$g(y)$ saturates at a constant value near resonances.

The chemical potential is studied by
evaluating $\mu=\partial E_T /\partial N_T$. In the limit of large
scattering length, the main
characteristic is that $\mu$ saturates at a value of around $80\%$ of the Fermi energy $\epsilon_F$
of the corresponding density. When the chemical potential in units of $\epsilon_F$
is plotted against scattering lengths $y=\frac{a}{d}$, all data again collapse to a single master curve
which quantitatively defines the scaling function $h(y)(=\frac{\mu}{\epsilon_F})$
introduced above; and $h(y)$ approaches
$0.8$ once $y$ becomes much bigger than unity (see
Fig.\ref{fig2}).
$v_s$, the speed of sound that depends on the compressibility of BECs,
can also be obtained by using the general relation $v_s^2= \rho_0/m (\partial \mu/\partial \rho_0)$.

\begin{figure}
\includegraphics[width=\columnwidth]{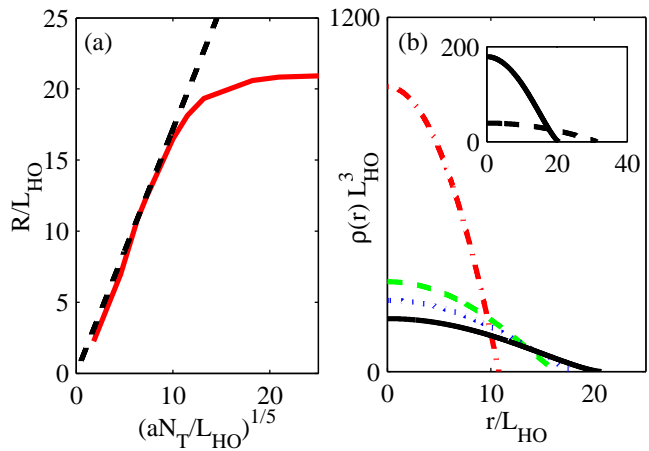}
\caption{(Color online) a) Radius $R$
of BECs in a spherical harmonic trap with harmonic length $L_{HO}$ as a
function of scattering length $a$; the dashed line is the mean field
Thomas-Fermi radius $R_{TF}$ (see discussions before Eq.(\ref{eqn:R}) ).
b)Spatial density profiles in a harmonic trap at different scattering
lengths; the density at the center $\rho(0)$ is estimated to be
$\rho(0)a^3=0.00012, 0.038, 0.24,180$
for the dashed-dotted, dashed, dotted, and solid line respectively.
In the inset, we also plot the mean field result(black dashed line)
for $\rho(0)a^3=180$ for a comparison.
In a),b), the total number of atoms is set to be $N_T=2\times
10^6$.}
\label{fig3}
\end{figure}

The scattering-length dependence of the chemical potential discussed here
implies a very peculiar evolution of sizes of BECs in a trap (with a
harmonic
length $L_{HO}$) when scattering lengths $a$ are increased.
In the limit of a small scattering length the size of condensates increases
as a function of scattering length $a$ and the {\em mean field}
Thomas-Fermi
radius in a spherical trap is $R_{TF}/L_{HO}=
(15 N_T a/L_{HO})^{1/5}$\cite{Dalfovo99}.
As the chemical potential saturates
at a value of $0. 8 \epsilon_F$ when scattering lengths become much
larger than the typical inter-particle
distance in a trap, the radius of the BECs
in this strongly interacting regime is also expected to approach a value
of
\begin{equation}
\frac{R}{L_{HO}}= A N_T^{1/6};\label{eqn:R}
\end{equation}
Numerical calculations further show that
$A=1.9$. As another application of our variational
approach, we quantitatively investigate radii of BECs near resonances
using a local density approximation (see Fig.\ref{fig3}).

In conclusion, we have examined basic properties of cold bosonic atoms
at large scattering lengths.
Using the variational method, we estimate various properties that can be potentially tested in
future cold atom experiments. Near resonances, we have found that
the chemical potential, speed of sound, and
the spatial density profile of cold bosons in a trap resemble the corresponding
properties of Fermi gases. This
particular aspect is also a unique feature of one-dimensional
Tonks-Girardeau gases where bosons are
viewed as fermionized particles\cite{Girardeau60,Paredes04,Kinoshita04}.
Our results are applicable near Feshbach resonances but before the Efimov physics fully sets in.
This work is supported by NSERC, the Canada and Canadian Institute for
Advanced Research. We thank Jason T. L. Ho for a stimulating discussion.

\end{document}